\documentclass[%
aps,
pre,
superscriptaddress,
notitlepage,
longbibliography,
10pt,
onecolumn
]{revtex4-1}
\usepackage[utf8]{inputenc}
\usepackage[T1]{fontenc}
\usepackage{amsmath, amssymb, amsfonts, bm}
\usepackage{graphicx}
\usepackage{color}
\usepackage{float}
\usepackage[colorlinks=true,urlcolor=black,linkcolor=blue,citecolor=blue]{hyperref}
\usepackage{subfigure}

\begin{document}

\title{Flow through three-dimensional self-affine fractures}

\author{H. J. Seybold} 
\affiliation{Departamento de F\'isica, Universidade Federal do Cear\'a,
    Campus do Pici, 60451-970 Fortaleza, Cear\'a, Brazil}
\affiliation{Physics of Environmental Systems, D-USYS, ETH, Zurich, 8093
	Zurich, Switzerland}

\author{H. A. Carmona} 
\affiliation{Departamento de F\'isica, Universidade Federal do Cear\'a,
	Campus do Pici, 60451-970 Fortaleza, Cear\'a, Brazil}

\author{F. A. Leandro Filho} 
\affiliation{Departamento de F\'isica, Universidade Federal do Cear\'a,
	Campus do Pici, 60451-970 Fortaleza, Cear\'a, Brazil}

\author{A. D. Ara\'ujo}
\affiliation{Departamento de F\'isica, Universidade Federal do Cear\'a,
	Campus do Pici, 60451-970 Fortaleza, Cear\'a, Brazil}

\author{F. Nepomuceno Filho}
\affiliation{Departamento de F\'isica, Universidade Federal do Cear\'a,
	Campus do Pici, 60451-970 Fortaleza, Cear\'a, Brazil}

\author{J. S. Andrade Jr.}
\email{soares@fisica.ufc.br}
\affiliation{Departamento de F\'isica, Universidade Federal do Cear\'a,
	Campus do Pici, 60451-970 Fortaleza, Cear\'a, Brazil}

\begin{abstract}
We investigate through numerical simulations of the Navier-Stokes equations the
influence of the surface roughness on the fluid flow through fracture joints.
Using the Hurst exponent $H$ to characterize the roughness of the self-affine
surfaces that constitute the fracture, our analysis reveal the important
interplay between geometry and inertia on the flow.  Precisely, for low values
of Reynolds numbers Re,  we use Darcy's law to quantify the hydraulic resistance
$G$ of the fracture and show that its dependence on $H$ can be explained in
terms of a simple geometrical model for the tortuosity $\tau$ of the channel. At
sufficiently high values of Re, when inertial effects become relevant, our
results reveal that nonlinear corrections up to third-order to Darcy's law are
aproximately proportional to $H$.  These results imply that the 
resistance $G$ to the flow follows a universal behavior by
simply rescaling it in terms of the fracture resistivity and using an effective
Reynolds number, namely, Re/$H$. Our results also reveal the presence of
quasi-one-dimensional channeling, even considering the absence of shear
displacement between upper and lower surfaces of the self-affine fracture.
\end{abstract}

\maketitle

\section{Introduction}

Understanding the behavior of a fluid flowing through a fractured rock is of
great importance in many practical applications~\cite{Sahimi1993, Berkowitz2002,
Sahimi2011, Osborn2011, Williams2017}.  In particular, it is crucial to
investigate how the fracture's surface morphology influences the flow resistance
in driving fluids through naturally or artificially fractured carbonate
reservoirs~\cite{Warren1963, Liu2016a, HUBBERT1957, Rubinstein2015}.  Since at
the reservoir scale fractures are mostly composed of networks of interconnected
cracks with very different sizes, it is important to understand how the behavior
of the flow in a single fracture scales with its size, as well as how it is
affected by the details of its geometry~\cite{Liu2016a}.  The local flow
structures are a direct result of the fracture's morphology and many studies
have been devoted to understand their upscale in order to derive consistent
macroscopic relations~\cite{Roux1993, Talon2010, Talon2010a, Wang2016}.

It is generally accepted that the morphology of brittle fractures follows
\emph{self-affine} scaling laws~\cite{Bunde1996}. More precisely, it means that
by re-scaling an in-plane vector $\mathbf{r}$ by $\lambda \mathbf{r}$, the
out-of-plane coordinate $z$ needs to be re-scaled by $\lambda^{H} z$ for the
surface to remain statistically invariant, where the scaling exponent $H$ is
called the Hurst exponent. It was first suggested that rock fractures have a
unique exponent $H=0.8$~\cite{Bouchaud1990, Maaloey1992, Schmittbuhl1993,
Cox1993,Schmittbuhl1995,Bouchaud1997, Oron1998}. However, recent studies
indicates that some natural fractured systems exhibit other values of $H$
(ranging from $0.45$ to $0.85$) depending on the material and the
fracturing process~\cite{Odling1994,Amitrano2002, Ponson2006,
Babadagli2015}. As a result, more than one universality class exists for
fractured rocks and therefore it is important to understand how the flow
properties are affected by variations in the Hurst exponent.

A single-phase flow in a fractured rock is usually characterized in terms of
Darcy's law~\cite{Sahimi1994,Sahimi2011, Dullien1992}, which defines a linear
relation between the mean flow velocity $ U $ and the pressure drop $\Delta P$
across the system, namely $U = -k \Delta P/\mu L$. Here $\mu$ is the fluid's
viscosity,  $L$ is the length of the fracture in the flow direction, and the
proportionally constant $k$ is the permeability. 
Essentially Darcy's law is a good approximation at low Reynolds numbers,
$\mathrm{Re} = \rho U w/\mu \ll 1$,
where $w$ is usually taken as the aperture of the fracture and $\rho$ is the
density of the fluid.  However, in order to understand the interplay between the
geometry and the flow inside the fracture, it is necessary to examine
\emph{local} aspects of the surface roughness and relate them to the relevant
mechanisms of momentum transfer through viscous and inertial forces.

The influence of surface roughness on the flow properties inside a fracture has
been first studied theoretically by Roux \emph{et al.}~\cite{Roux1993}.  They
predicted that the permeability of a self-affine fracture should scale with the
length of the system as $k \sim L^{2H}$. This result is based on the assumption
that the fracture behaves like a system of parallel plates with an effective
aperture $w$, where the self-affinity implies that $w \sim L^H$, and $k \sim
w^2$ follows the solution of the Stokes equation.
Since then, several theoretical and experimental studies focused on how the
permeability scales with the fracture's opening and length in the viscous flow
regime. Using perturbation theory, Drazer and Koplik~\cite{Drazer2000}
calculated for two-dimensional flows that the permeability should scale as $k_0
- k \sim L^H$, where $k_0$ is the permeability of the unperturbed system. Later
they extended their study to three-dimensional flows and confirmed these results
in terms of the effective medium analysis and numerical simulations at low
Reynolds numbers.
Talon \emph{et al.}~\cite{Talon2010} conjectured that the permeability is
controlled by the minimum aperture of the fracture  $w_{min}$ and found that, for
two-dimensional flows, $k\propto w_{min}^{3-1/H}$.
For three-dimensional flows, however, Talon \emph{et al.} have shown
numerically that $k \sim  w_{min}^{2.25}$ for $H=0.8$ and $k \sim  w_{min}^{2.16}$ for
$H=0.3$.

The role of inertia on fluid flow through two-dimensional self-affine fractures
has been addressed by Sketne \emph{et al.}~\cite{Skjetne1999}, who considered
fractures with constant aperture and $H=0.8$. Their numerical simulations show
that, in the range of intermediate Reynolds number $\mathrm{Re} \approx 1$,
the flow can be described by a weak inertia equation~\cite{Mei1991, WODIE1991},
whereas for moderate Reynolds numbers ($25 \leq \mathrm{Re} \leq 52$) 
inertial effects can be described by the Forchheimer
equation~\cite{Forchheimer1901}. More recent numerical studies~\cite{Briggs2017}
extended these results for different Hurst exponents with long range correlations,
namely $H>0.65$.

It is evident that inertia has very different impact on two- and
three-dimensional flow systems. Here we address the question on how the
permeability and the nonlinear corrections to Darcy's law depend on the surface
roughness of three-dimensional self-affine fractures. To do so we
systematically examine the behavior of the fluids hydraulic resistance as a
function of the Reynolds number in the range from $\mathrm{Re}=10^{-2}$ to
$\mathrm{Re}=500$, and for different values of the Hurst exponent, varying from
strongly anticorrelated $H=0.3$ to strongly correlated values, $H=0.9$.

The remainder of this paper is organized as follows. Section~\ref{sec:methods}
describes the methodology we have used to generate the geometry, and the setup of
the computer simulations. In Section~\ref{sec:results} we present and discuss our
simulation results and Section~\ref{conclusion} is devoted to the conclusions.

\section{Methods}\label{sec:methods}

\begin{figure} [!ht]
    \centering
    \includegraphics[width=\textwidth]{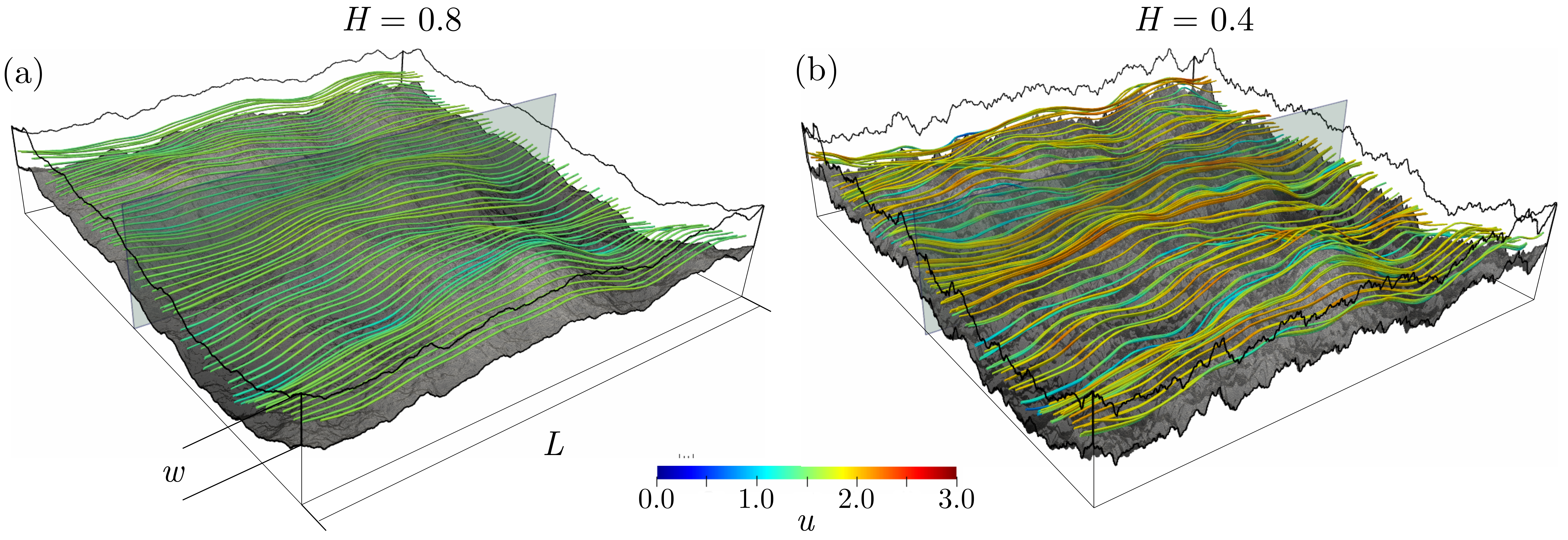} %
    \caption{           
        Fluid flow through a typical fracture joint calculated for
        $\mathrm{Re}=100$ and Hurst exponents (a) $H=0.8$ and (b) $H=0.4$.
        The fluid flows from left to right. Streamlines are
        also shown, colored according to the local velocity magnitude.        
    } \label{fig1}
\end{figure}

The three-dimensional numerical domain used in our analysis consists of the
volume between  two identical self-affine surfaces, representing the fracture
walls. The surfaces have been displaced perpendicular to the
mean surface plane (the $x-y$ plane). Specifically, no additional shear
displacement is added to the surfaces in this plane, so that the fracture
aperture $w$ is constant throughout the numerical domain (see Fig.~\ref{fig1}).

The wall surfaces are generated using a two-dimensional generalization of
the fractional Brownian Motion~\cite{Oliveira2011, Morais2011, Mandelbrot1968,
Peitgen2011} which satisfies the following scaling relation:
\begin{equation}\label{eq_scaling}
    \left<\left[
        z\left( \mathbf{r}_2 \right) - z\left( \mathbf{r}_1 \right)			
    \right]^2 \right> = \sigma_{z}^2
    \left|    
    \frac{ \mathbf{r}_2 - \mathbf{r}_1}{L}
    \right|^{2H},
\end{equation}
where $\sigma_z^2$ is the mean square increment for points separated by a
distance $L$. Here $z(\mathbf{r})$ defines the elevation of the wall surface and
$\mathbf{r}$ is a vector in the $x-y$ plane. The Hurst exponent $H$
characterizes the spatial correlations of the surface. Surfaces with $H<0.5$ are
spatially anti-correlated, while for $H>0.5$ long-range spatial correlations are
present. For the case $H=0.5$ we obtain ordinary Brownian surfaces formed by
successive uncorrelated increments~\cite{Peitgen2011}. In order to create a
discrete fractional Brownian surface with a given exponent $H$ numerically,
we use the Fourier filtering method~\cite{Earnshaw1991, Peitgen2011}. This
method imposes a scaling behavior on the spectral density $S_z$ as,
\begin{equation}\label{Eq_specral_dens}
    S_z(k) \propto \frac{1}{k^\xi},
\end{equation}
where the parameter $\xi$ is related to the Hurst exponent via $\xi=2+2H$ for
two-dimensional surfaces~\cite{Hansen2001}. Equation~\eqref{Eq_specral_dens} is
used to define the amplitudes of the discrete Fourier
spectrum of the wall surface, which is then transformed  back to real
space using a fast Fourier transform.

For all fracture realizations, the aperture is kept constant, $w=5$, and the
length in $x$ and $y$ directions is set to be $L=500$, both in dimensionless
units. In order to obtain fractures with a comparable variability in the
$z$-direction, the amplitude of the surfaces is also fixed for all realizations
to be $\sigma_{z}=5$ in dimensionless units.

We now turn our attention to the flow between the two rough fracture walls.
The three-dimensional flow is described by the
incompressible Navier-Stokes equations under isothermal steady-state conditions.
The momentum and mass conservation equations are written as
\begin{gather}
    \rho~\mathbf{u\cdot\nabla u} = -\mathbf{\nabla} p +
    \mu~\mathbf{ \nabla }^{2} \mathbf{u}~, \label{Eq_momentum} \\
    {\bf{\nabla\cdot u}}=0~, \label{Eq_continuity}
\end{gather}
where $\mathbf{u}$, $p$ and  $\rho$ are the velocity, pressure, and the fluid's
density, respectively. We apply non-slip boundary conditions at the top and
bottom walls. The fluid is injected in the $x$ direction at $x=0$ using a
uniform velocity profile with amplitude $U$ at the inlet, and a constant
pressure defines the outlet boundary, at $x=L$. Laterally symmetrical boundary
conditions were applied to minimize finite-size effects. 
In order to solve Eqs.~\eqref{Eq_momentum} and~\eqref{Eq_continuity}
numerically, we first discretize the volume between the top and bottom surface
of the fracture using a tree-dimensional unstructured hexahedral mesh generated
using the OpenFOAM's meshing tool \emph{snappyHexMesh}~\cite{OF_Weller1998}. 
Close to the surface, the hexahedral cells were refined three times in order to
capture small variations of the fracture surface.

For each value of the Hurst exponent in the range $0.3 \le H \le 0.9$, we
generated five realizations of the computational domain to compute ensemble
averages.  For each realization, flow simulations are performed with different
values of the Reynolds numbers in the range $0.01 \leq \mathrm{Re} \leq 500$ by
adjusting the inlet velocity $U$.

\section{Results and Discussion}\label{sec:results}

\begin{figure}[!h]
    \centering
    \includegraphics[width=112.2mm]{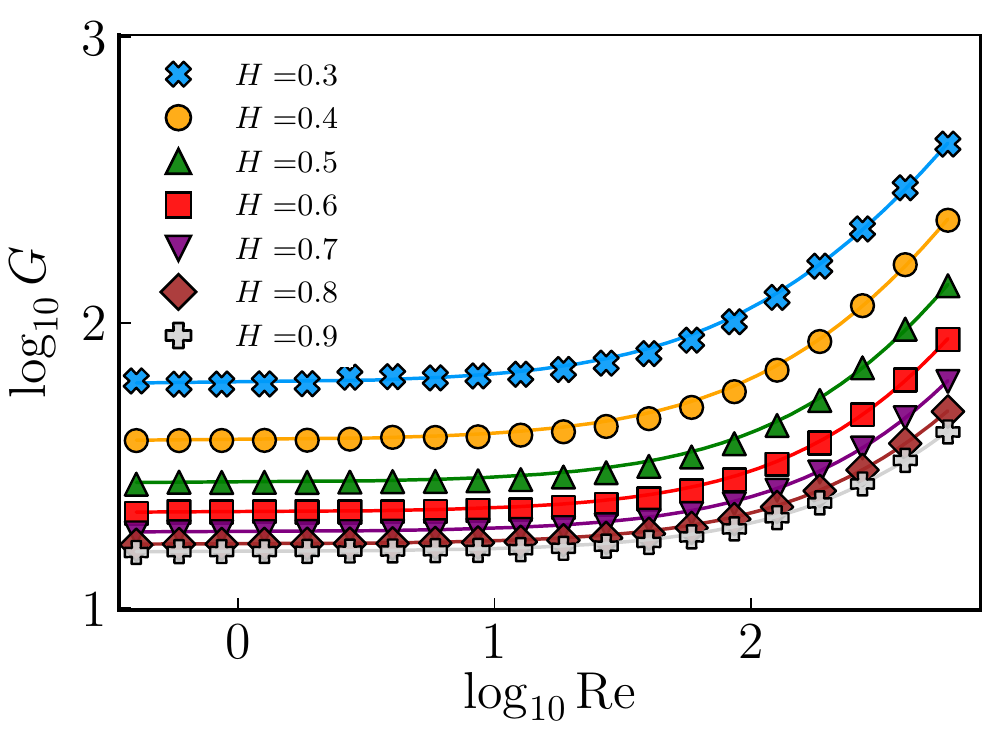} %
    \caption{          
        Hydraulic resistance $G$ as a function of the Reynolds number $\mathrm{Re}$
        for different values of the Hurst exponent $H$. In all cases, the plateau
        corresponding to Darcy's law (constant $G$) is followed by a non-linear regime
        that reflects the effect of convection on the flow. The error bars are smaller
        than the symbols and the solid lines are the best fit to the data using
        Eq.~\eqref{Eq_conductance}.  For each value of the parameters $\mathrm{Re}$ and
        $H$, the value of $G$ is obtained as the average over a total of five
        realizations.
        } \label{fig2}
\end{figure}

The Forchheimer equation~\cite{Forchheimer1901, Whitaker1996} has been
extensively used as an extension of Darcy's law to account for inertial
corrections in flow through disordered pore
structures~\cite{Sahimi2000,Sahimi2011,Dullien1992}.  Expectedly, the addition
of higher-order corrections in the velocity to the Forchheimer equation allows
for a better agreement with experimental data over the full range of the laminar
regime ~\cite{Sahimi1994, Edwards1990, Andrade1999, Hill2001}. Up to cubic
order, these corrections can be written as,
\begin{equation}
-\frac{\Delta P}{L} = \alpha \mu U + \beta \rho {U^2} + \frac{\gamma \rho^{2}
    U^{3}}{\mu}~,
\label{Eq_cubic}
\end{equation}
where $\alpha \equiv 1/\kappa$ corresponds to the reciprocal of the permeability
of the channel, and $\beta$  and $\gamma$  are the coefficients of the second-
and third-order corrections, respectively.  Rewriting Eq.\eqref{Eq_cubic} in
terms of $\mathrm{Re}$, we obtain,
\begin{equation}\label{Eq_conductance}
    G = \alpha w^2 + \beta w \mathrm{Re} + \gamma \mathrm{Re}^{2},
\end{equation}
where $G \equiv -\Delta P w^{2}/ \mu U L$ is a dimensionless measure of the
hydraulic {\it resistance} of the fracture. Figure~\ref{fig2} displays the
results from all our numerical simulations, where $G$ is plotted as a function
of Reynolds number and for different values of the Hurst exponent.  The solid
lines are the nonlinear fits of Eq.~\eqref{Eq_conductance} to the data sets in
order to determine the coefficients  $\alpha$, $\beta$ and $\gamma$.

\begin{figure}
    \centering
    \includegraphics[width=112.2mm]{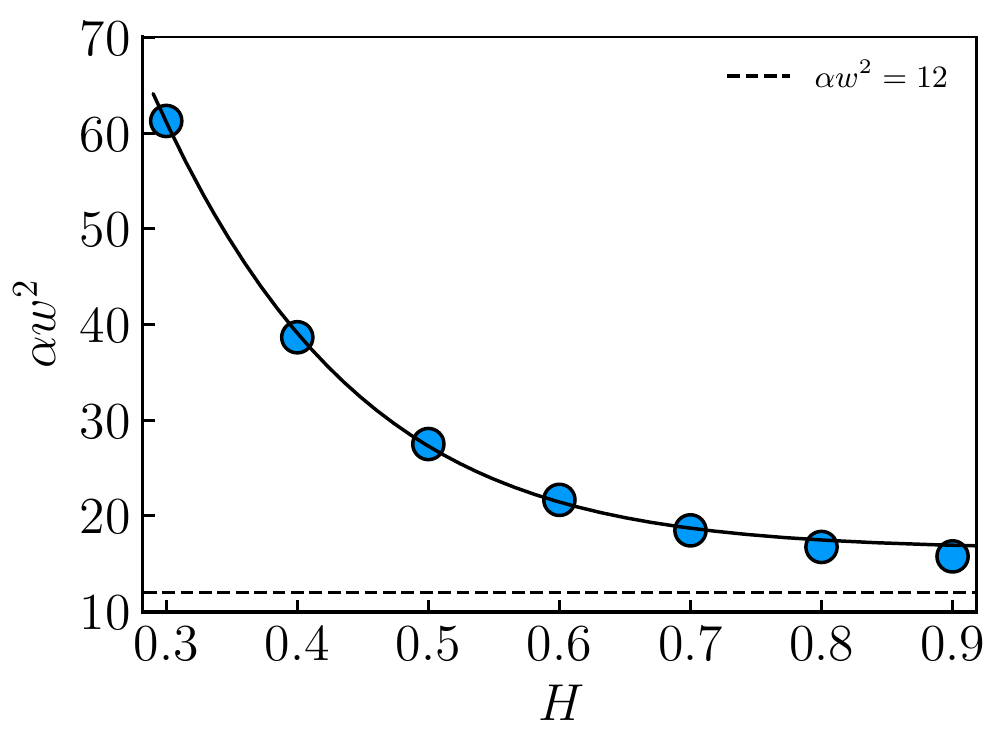}
    \caption{ Dependence of $\alpha w^2$ on the Hurst exponent $H$. The parameter
$\alpha\equiv 1/\kappa$ aproaches the limiting value of $12/w^2$  (dashed black
line), as expected for the Poiseuille flow.  The solid black line is obtained by
the combination of Eq.~\eqref{eq_alphaw2}, with the previously estimated parameters $a$ and
$b$, and  Eq.~\eqref{eq_tau}, with $\delta
x = 2$, $\sigma_z = 5$ and $L=500$.
}\label{fig3}
\end{figure}

\begin{figure} %
    \centering %
    \includegraphics[width=112.2mm]{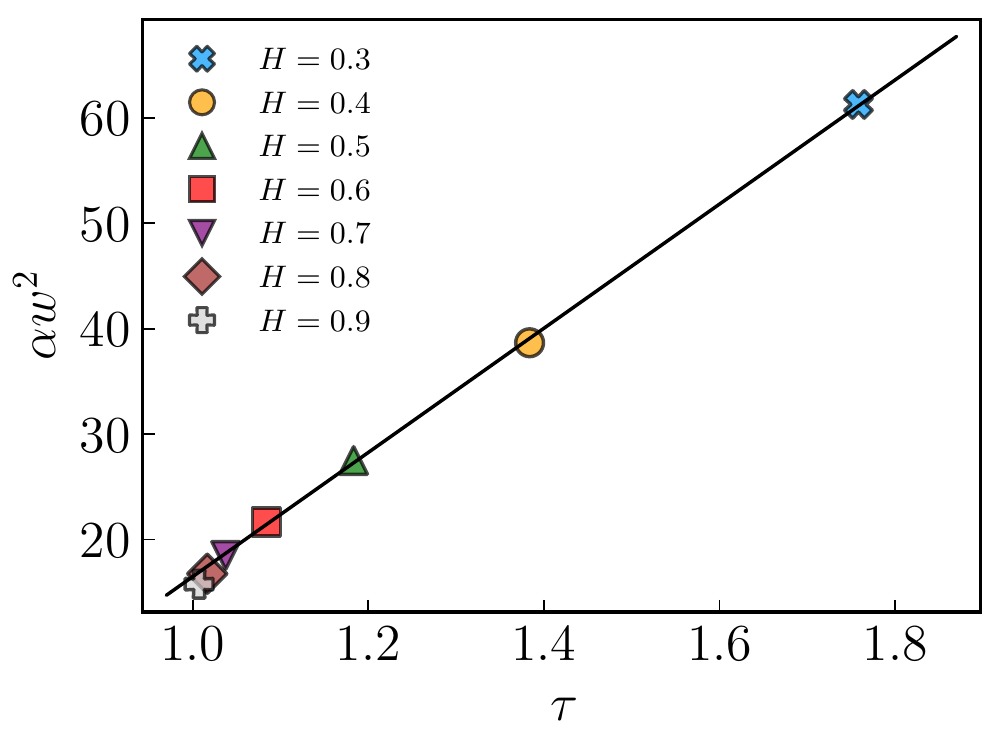}
    \caption{
        Variation of $\alpha w^2$ with the tortuosity $\tau$. The error
        bars are smaller than the symbols. The solid
        lines correspond to the least-squares fit to the simulation data using $\alpha w^2=a + b
        \tau$, with $a=-42.4\pm0.7$ and $b=58.9\pm0.6$, where $\tau$ was
        computed from Eq.~\eqref{eq_tau} with $\sigma_z=5$ and $\delta = 2$.
    }\label{fig4} %
\end{figure}

For small Reynolds numbers, $G$ is dominated by the viscous term in
Eq.~\ref{Eq_conductance}, namely, $\alpha w^2$, which decreases monotonically
with the Hurst exponent $H$ as shown in Fig.~\ref{fig3}.  Consistent with the
results for a Poiseuille flow between two parallel planes with constant aperture
$w$, it approaches the value $\alpha = 12/w^2$ for large values of $H$.  In
order to understand the particular form of this relation, we consider, as a
first approximation, the fracture as composed of a sequence of parallel plates
with varying angles with respect to the $x-y$ plane. 
For a self-affine fracture surface, one can define a
\emph{tortuosity} factor as, 
\begin{equation}\label{eq_tau}
    \tau \equiv  \frac{L_{p}}{L} = 
    \sqrt{1 + \left(\frac{\sigma_z}{\delta x}\right)^2 \left(\frac{L}{\delta x}\right)^{-2H}},
\end{equation}
where $L_{p}$ is the perimeter in the direction of the flow and $\delta x \ll L$ is
the numerical resolution used to generate the rough surface (see Section I of
the Supplemental Material).
Considering this simplified geometrical model, we now conjecture that the first
term in Eq.~\eqref{Eq_conductance} can be described as, 
\begin{equation}\label{eq_alphaw2}
    \alpha w^{2} = a\tau + b.
\end{equation}
Here, to  be consistent with the limiting value for the hydraulic resistance of
parallel plates at very low Re, the parameters $a$ and $b$ should be obtained
from the numerical simulations with the constraint that $a+b=(\alpha
w^{2})_{min} \geq 12$.  Figure~\ref{fig4} shows that $\alpha w^2$ follows very
closely Eq.~\eqref{eq_alphaw2} for $H \leq 0.7$.  From the least-squares fit to
the data, combining Eqs.~\eqref{eq_tau} and \eqref{eq_alphaw2} with
$\sigma_{z}=5$ and $\delta x=2$, we obtain $a=-42.4\pm0.7$ and $b=58.9\pm0.6$,
which consistently gives $a+b=16.5 \geq 12$.  For large values of the Hurst
exponent, the influence of the self-affine geometry on the flow is less strong
because the local slopes in the channel are not so high as compared to those
present in channels generated with smaller values of $H$. As a consequence, this
purely geometrical model tends to overestimate the hydraulic resistance $\alpha
w^{2}$ for $ H > 0.7$.  This systematic discrepancy is better visualized in
Fig.~\ref{fig3}, where the solid black line is also obtained from Eqs.~(7) and (8), with
the same previously estimated parameters $a$ and $b$. 

%

\begin{figure} [!h] %
    \centering %
    \includegraphics[width=112.2mm]{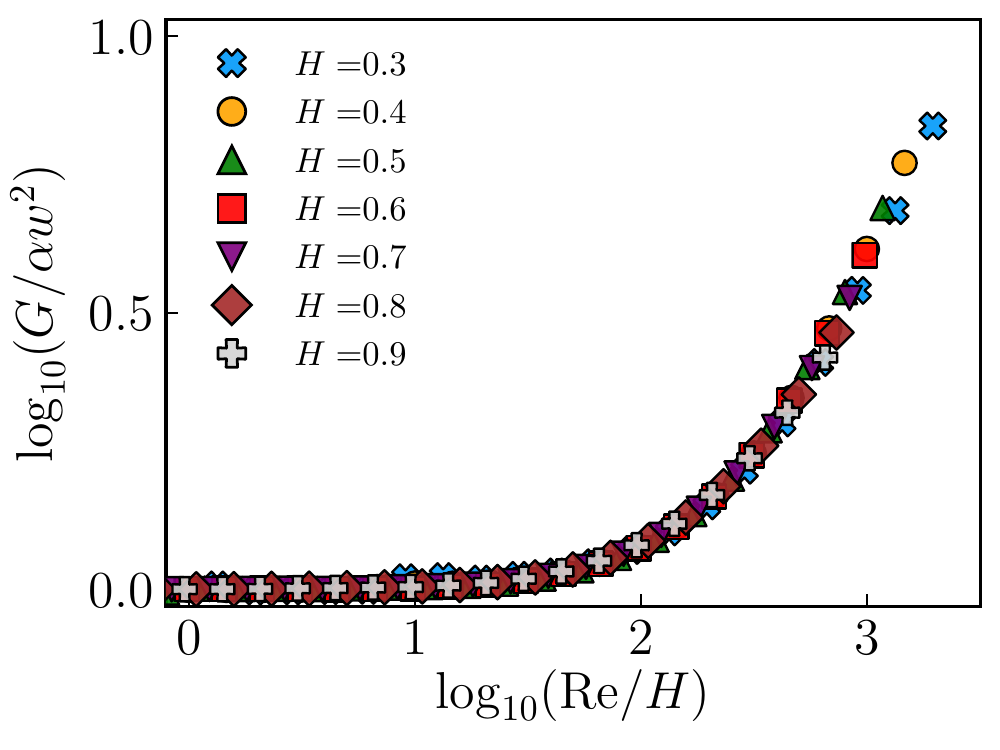}
    \caption{
        Hydraulic resistance $G$ as a function of $\mathrm{Re}/H$. The excellent
        collapse of the simulation data onto a single curve indicates that the onset of
        the nonlinear contributions to the Darcy's law increases 
        linearly with the Hurst exponent.
    } \label{fig5} %
\end{figure}

\begin{figure}[!h] %
    \centering %
    \includegraphics[width=\textwidth]{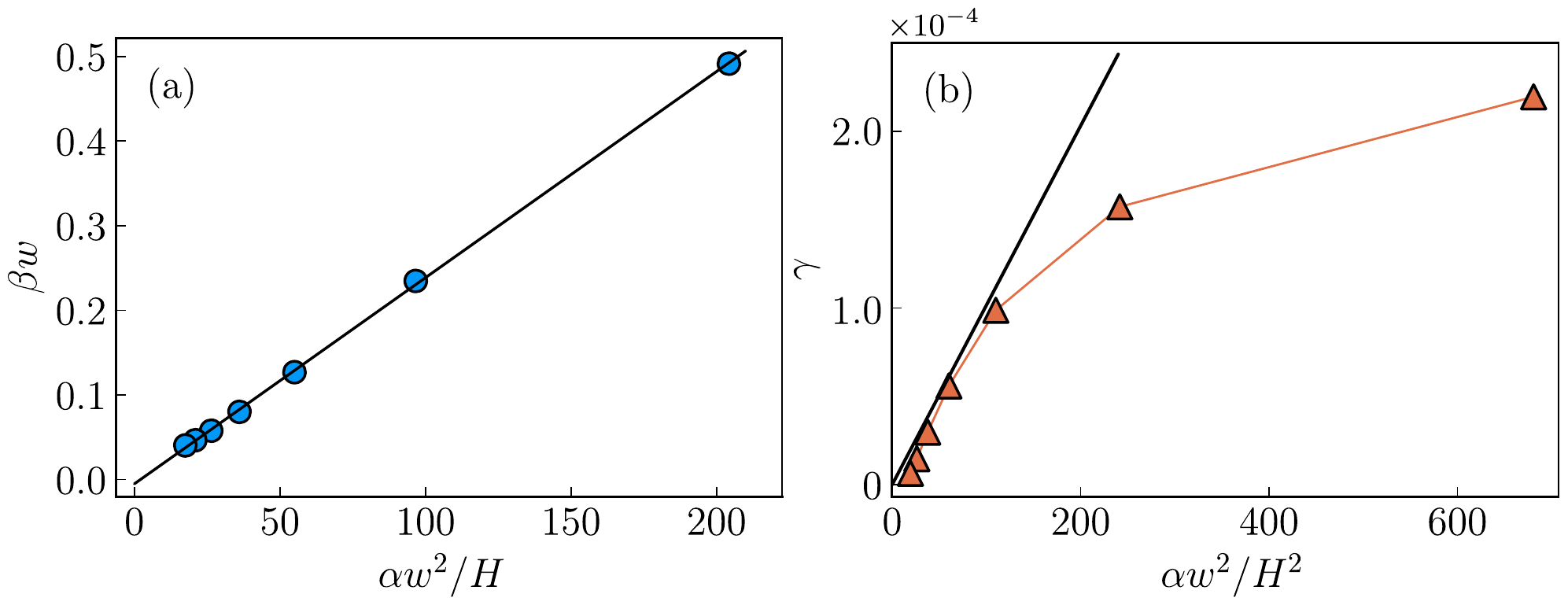}
    \caption{
        (a) The parameter $\beta w$ as a function of $ \alpha w^2/H$. The black
        solid line corresponds to the least-squares fit to the data using
        $\beta w = c_0 \left(\alpha w^2/H\right) $ with $ c_0=0.0023 \pm 0.0001$. 
        (b) The scaling of $\gamma$ with $\alpha w^2/H^2$. The
        solid line corresponds to $ \gamma = c_1 \left(\alpha
        w^2/H^2\right)$ with $c_1 = (1.01\pm0.05)\times10^{-6}$. The simulation data deviates
        from the expected linearity for small $H<0.4$ ($\alpha w^2/H^2 > 110$).}\label{fig6} %
\end{figure}

In agreement with experiments~\cite{Chen2017a}, we observe that the transition
from a linear (constant $G$) to a non-linear regime occurs at lower
$\mathrm{Re}$ for rougher surfaces. Although the absolute value of $G$
significantly depends on the tortuosity, and decreases with $H$, our results
shown in Fig.~\ref{fig2} suggest that the general increasing trend of the nonlinear
corrections as a function of $\mathrm{Re}$ is independent of the Hurst
exponent. In order to quantify the impact of the surface roughness on the
departure from Darcy's law, we plot $G$  as a function 
of an effective Reynolds number, defined as $ \mathrm{Re}/H $.
Following this procedure, all curves collapse onto a single master curve, as
shown in Fig.~\ref{fig5}.  This collapse is an indication that the onset of the
non-linear contributions to the hydraulic resistance increases in a linear
fashion with the parameter $H$. As a matter of fact, the excellent quality of
the collapse implies the scaling relations,
\begin{equation}
    \beta w \propto \alpha w^2/H
\end{equation}\
and
\begin{equation}
    \gamma \propto \alpha w^2/H^2.
\end{equation}
As depicted in Fig.~\ref{fig6}, the second-order term follows the proposed
scaling relations rather well. For the third-order coefficient, however, we
observe significant deviations from the proposed linear trend for $H<0.4$
($\alpha w^2/H^2 > 110$).

\begin{figure} [!ht]
    \centering
    \includegraphics[width=112.2mm]{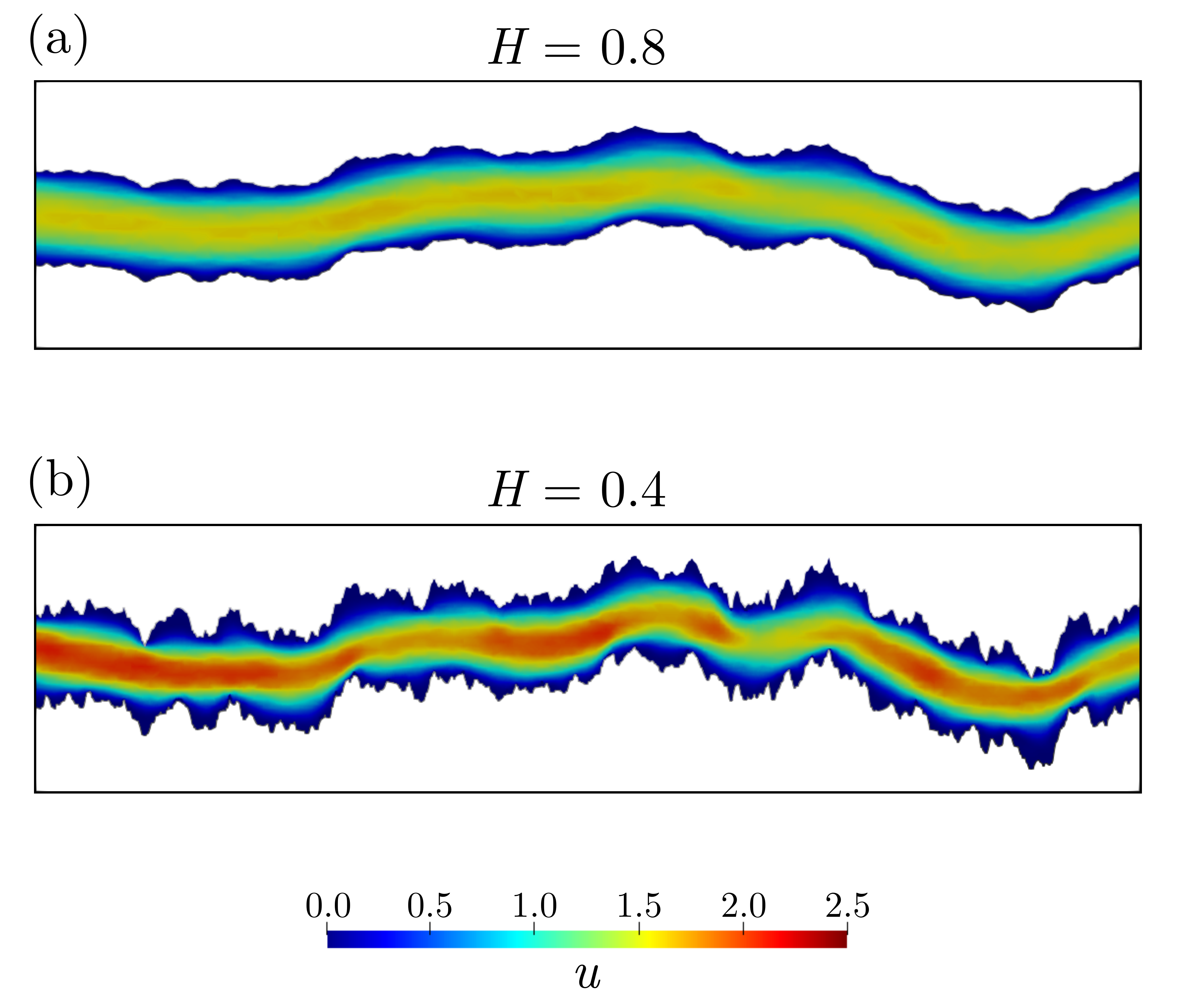} %
    \caption{  
        Contour plots of the velocity magnitudes in the cross sections along the
        channels as indicated in Fig.~\ref{fig1}, for $\mathrm{Re}=100$ and Hurst exponents
        (a) $H=0.8$ and (b) $H=0.4$.          
        } \label{fig7}
\end{figure}

Next, we analyze the impact of the surface roughness on the velocity field
directly. Figures~\ref{fig1}a and ~\ref{fig1}b show typical realizations of
fluid flows through two fracture joints at $\mathrm{Re} = 100$ for Hurst
exponents $H = 0.4$ and $0.8$, respectively. Streamlines of the velocity field
are also shown, being colored by the velocity magnitude.  The contour plots in
Fig.~\ref{fig7} denote the magnitude of the velocity field on the
cross-sectional planes along the main flow direction, as indicated in
Fig.~\ref{fig1}.  For the larger Hurst exponent, $H = 0.8$, the fluctuations in
the local velocity field are visually smoother, with the maximum velocity near
the center of the channel approximately equal to $3U/2$, as expected in the
limiting case of parallel plates~\cite{Batchelor2000} (see Section II of the
Supplemental Material).
For $H=0.4$, however, the situation is rather different. Due to continuity,
regions of higher velocities are clearly more confined at the center of the
channel, since the zones of almost stagnated flow close to walls broadens.  As a
consequence, the velocity magnitudes climb up to  $2.5U$, which is consistent
with a relatively higher effective Reynolds number, Re/$H$, as suggested by the
collapse in Fig.~\ref{fig5}. An enhancement in the velocity magnitude due to
local disorder in the surface morphology can then persist and propagate further
into the fracture joint forming preferential flow paths, where the fluid follows
trajectories connecting ``valleys'' and around ``mountains'' of the rough
surfaces. This effect is only possible in three-dimensional flows, since in two
dimensions the flow is forced through local bottlenecks.

A similar preferential channeling  effect has been found in previous
experiments~\cite{Ishibashi2015} and computational simulations~\cite{Drazer2002,
Lo2014, Huang2017}. This effect, however, has always been associated to an
additional shear displacement between the upper and lower surface, which
generates a heterogeneous aperture distribution throughout those fractures. Our
simulations, however, show that no lateral displacement is needed and that the
preferential channeling  effect thus should be a result of an effective aperture
field which is significantly affected by the surface topology.

The effect of preferential channeling can be quantified by the participation
ratio $\pi$, which has been previously utilized to describe the spatial
localization of kinetic energy inside the flow through disordered porous
media~\cite{Andrade1999}, being defined as,
\begin{equation}
    \pi = \frac{\left<e\right>^2}{\left<e^2\right>},
\end{equation}
where $ \left<e^n\right> = (1/V) \iiint
\left(\mathbf{u}\cdot\mathbf{u}\right)^{n} \; d^3\mathbf{r} $ is the $ n^{th} $
moment of the kinetic energy and $ V $ is the volume of the system.  If the
kinetic energy is uniformly distributed across the sample, one obtains $\pi
\rightarrow 1$, whereas if the flow field is strongly localized, $ \pi
\rightarrow 1/V$, approaching zero in the limit of an infinitely large system.

\begin{figure}[!h]
        \includegraphics[width=112.2mm]{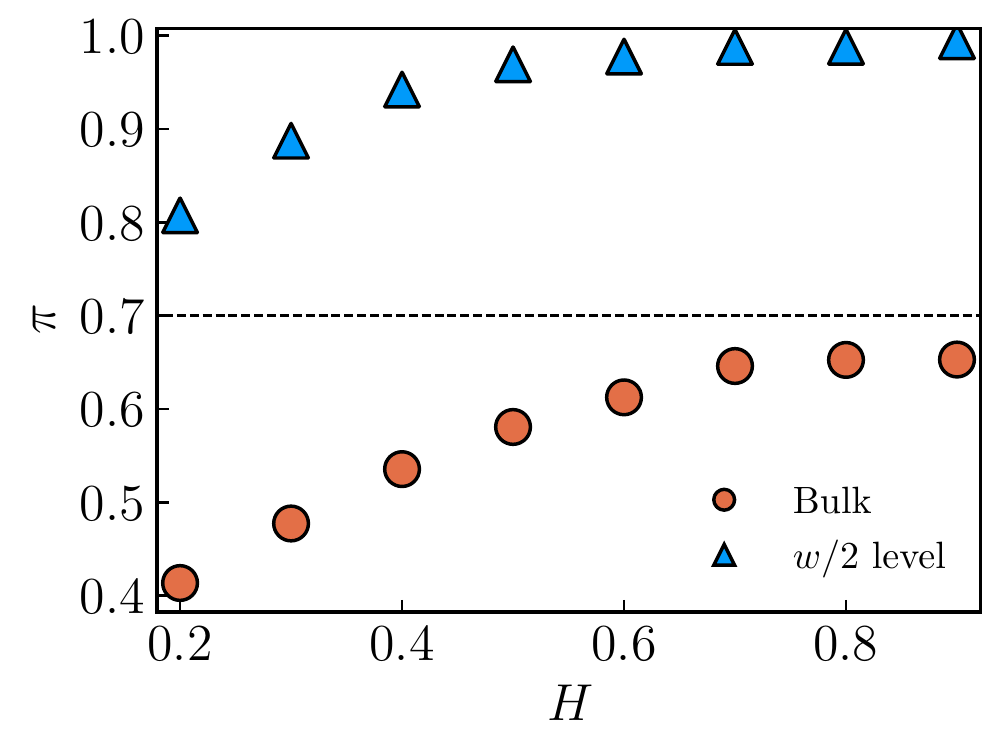}
        \caption{
            Participation index $\pi$ as a function of the Hurst exponent $H$.
            Red circles correspond to the bulk participation index and the blue
            triangles denote the participation index in the $w/2$-level surface,
            obtained by translating the bottom surface of the channel by $w/2$
            in the $ z $-direction. The dashed horizontal line corresponds to
            $\pi=\pi_{0}$, namely, the value of participation for the flow at low
            $\mathrm{Re}$ between two parallel plates. The results were obtained
            by statistically averaging over five realizations.
        } \label{fig8} %
    \end{figure}

Figure~\ref{fig8} shows the variation of $\pi$ as a function of $H$ for
$\mathrm{Re} = 100$. For $H\rightarrow 1$ the participation ratio approaches a
value of $\pi_0=0.7$, which is the expected value for a Poiseuille flow between
two parallel plates (see Section III of the Supplemental Material).
By decreasing $H$, the participation ratio decreases monotonically indicating a
stronger preferential channeling effect. Also shown in Fig.~\ref{fig8}  is the
participation ratio computed only in the  $ w/2 $-level surface, defined as the
surface for which all points are at the vertical half distance between the lower
and upper boundary of the crack.  Compared to the bulk flow, the kinetic energy
is distributed more homogeneously in this surface as the effect of the wall
roughness is minimal. In this case, $\pi$ also increases with the Hurst
exponent, reflecting the formation of flow channels in fractures generated with
low values of $H$.

\section{Conclusions}\label{conclusion}

In summary, we have presented an extensive numerical study of single-phase flow
through three-dimensional self-affine fracture joints. 
Our results show that the hydraulic resistance of the fracture to the flow $G$
at low Reynolds numbers follows Darcy's law with a dependence on the Hurst
exponent that can be explained in terms of a purely geometrical model for the
channel tortuosity $\tau$. For higher values of Re, at the onset of inertial
effects in the flow, we find that nonlinear second- and third-order corrections
to Darcy's law are approximately proportional to $H$. These results enable us to
propose a universal curve to describe the variation of $G$ with Re at laminar
flow conditions and for any value of $H$.
Finally, we find that preferred flow paths arise in the flow field,
indicating that, even in three-dimensional fracture joints with no shear
displacement between top and bottom surfaces, the effective fracture aperture
field is heterogeneous.

\begin{acknowledgments}
	We thank the Brazilian agencies CNPq, CAPES and FUNCAP, also the National
	Institute of Science and Technology for Complex Systems and Petrobras for
	financial support.
\end{acknowledgments}


%
    
\end{document}